\documentclass[10pt,journal,compsoc,final]{IEEEtran}
\ifCLASSOPTIONcompsoc
  \usepackage[nocompress]{cite}
\else
  \usepackage{cite}
\fi
\ifCLASSINFOpdf
\else
\fi
\usepackage{amsmath}
\usepackage{subfig}
\usepackage{xcolor}
\usepackage{amsfonts}
\newcommand{\ineq}[1]{\footnotesize$#1$\normalsize}{}
\usepackage{gensymb}
\usepackage{graphicx,pifont}
\let\oldding\ding
\renewcommand{\ding}[2][1]{\scalebox{#1}{\oldding{#2}}}
\usepackage{threeparttable}
\usepackage{footnote}
\usepackage{subfig}
\makesavenoteenv{tabular}
\usepackage{tablefootnote}
\usepackage[hang]{footmisc}
\usepackage{multirow}
\usepackage{lipsum}

{}
\newcommand{\mr}[1]{\textcolor{black}{#1}}

\begin{document}
%
\title{A Framework to Explore Workload-Specific Performance and Lifetime Trade-offs in Neuromorphic Computing}
%
%
%
%

\author{Adarsha~Balaji, 
        Shihao~Song, 
        Anup~Das, 
        Nikil~Dutt, 
        Jeff~Krichmar,\\ 
        Nagarajan~Kandasamy, 
        Francky~Catthoor
\IEEEcompsocitemizethanks{\IEEEcompsocthanksitem A. Balaji, S. Song, A. Das, N. Kandasamy are with 
Drexel University, Philadelphia, PA, USA
E-mail:anup.das@drexel.edu.
\IEEEcompsocthanksitem N. Dutt and J. Krichmar are with the Department
of Computer Science, University of California, Irvine, CA, USA.
\IEEEcompsocthanksitem F. Catthoor is with Imec, Belgium and KU Leuven, Belgium.}
\thanks{Manuscript received Month DD, YYYY; revised Month DD, YYYY.}}
\markboth{IEEE Computer Architecture Letters,~Vol.~XX, No.~Y, Month~YYYY}%
{Balaji \MakeLowercase{\textit{et al.}}: A Framework to Explore Workload-Specific Performance and Lifetime Trade-offs in Neuromorphic Computing}
%



\IEEEtitleabstractindextext{%
\begin{abstract}
 Neuromorphic hardware with non-volatile memory (NVM) can implement machine learning workload in an energy-efficient manner.
Unfortunately, certain NVMs such as phase change memory (PCM) require high voltages for correct operation. These voltages are supplied from an on-chip charge pump. 
If the charge pump is activated too frequently, its internal CMOS devices do not recover from stress, accelerating their aging and leading to negative bias temperature instability (NBTI) generated defects.
Forcefully discharging the stressed charge pump can lower the aging rate of its CMOS devices, but makes the neuromorphic hardware unavailable to perform computations while its charge pump is being discharged.
This negatively impacts performance such as latency and accuracy of the machine learning workload being executed.
In this paper, we propose a novel framework to exploit workload-specific performance and lifetime trade-offs in neuromorphic computing. 
Our framework first extracts the precise times at which a charge pump in the hardware is activated to support neural computations within a workload.
This timing information is then used with a characterized NBTI reliability model to estimate the charge pump's aging during the workload execution.
We use our framework to evaluate workload-specific performance and reliability impacts of using 1) different SNN mapping strategies and 2) different charge pump discharge strategies. We show that our framework can be used by system designers to explore performance and reliability trade-offs early in the design of neuromorphic hardware such that appropriate reliability-oriented design margins can be set.
\end{abstract}

}

\maketitle

\IEEEdisplaynontitleabstractindextext

%
\IEEEpeerreviewmaketitle

\vspace{-40pt}
\section{Introduction}\label{sec:introduction}
\IEEEPARstart{A}{} neuromorphic 
hardware consists of artificial neurons and synapses to implement spiking neural networks (SNNs) \cite{Maass1997NetworksModels}. Emerging non-volatile memory (NVM) cells organized into crossbars are used to store synaptic strengths. Certain NVMs such as phase-change memory (PCM) require high voltages (\ineq{\sim 3V - 5V}) to read and program synaptic strengths. These high voltages not only create reliability issues for NVM cells in a crossbar, but also for the internal CMOS devices of the on-chip charge pump \cite{Shen2018ZeroRange}, which generates these voltages. In this paper, we study one specific high voltage related reliability issue of a charge pump in the context of neuromorphic computing -- that of threshold voltage (\ineq{V_\text{th}}) stress. 
If the charge pump is activated too frequently, its CMOS devices do not recover from stress, accelerating their \textit{aging} and eventually leading to failures.

Typically, a charge pump is several orders of magnitude larger than the size of a crossbar \cite{Shen2018ZeroRange}. To mitigate this large size, system designers connect many crossbars to each charge pump. 
Therefore, charge pump failures are a critical bottleneck to the prolonged operation of a neuromorphic hardware. Redundant charge pumps can improve reliability but increases hardware area. 
To improve reliability, stressed charge pumps can also be forcefully discharged, where a discharge operation involves applying a low voltage to all CMOS devices in the charge pump. Once discharged, the charge pump requires several cycles to boost its voltage back, before it can safely be used to access NVM cells in a crossbar. 
During this interval, crossbars are unable to process spikes, introducing a spike propagation delay. This delay negatively impacts performance (such as latency and accuracy) of the SNN workload being executed \cite{balaji2019mapping}.

Aging of a charge pump depends on how frequently NVM cells in the hardware are activated, which is due to spikes generated by the SNN workload being executed.

We propose a novel framework that allows system designers to explore workload-specific trade-offs involving reliability, performance, and design cost, early in the design process such that appropriate reliability-oriented design margins can be set.
Our framework incorporates the CARLsim simulator \cite{Chou2018CARLsim4} to first extract the precise times of spikes in a SNN workload. 
We then use a characterized reliability model to estimate aging of charge pumps based on their activation times, which are influenced by the mapping of synapses to crossbars and the connectivity of crossbars to charge pumps in the hardware. 
We show that this framework can be integrated inside 
1) {design-time techniques}, where neurons and synapses can be efficiently allocated to different crossbars, balancing aging of all charge-pumps, 2) {run-time techniques}, where stressed charge pumps can be forcefully discharged at appropriate intervals, minimizing their aging without significantly hurting performance, and 3) 
{architectural techniques}, where the number of charge pumps can be budgeted to achieve a target lifetime.

\vspace{-10pt}
\section{Background and Motivation}\label{sec:background}
\begin{figure}[t!]
	\centering
	\centerline{\includegraphics[width=0.99\columnwidth]{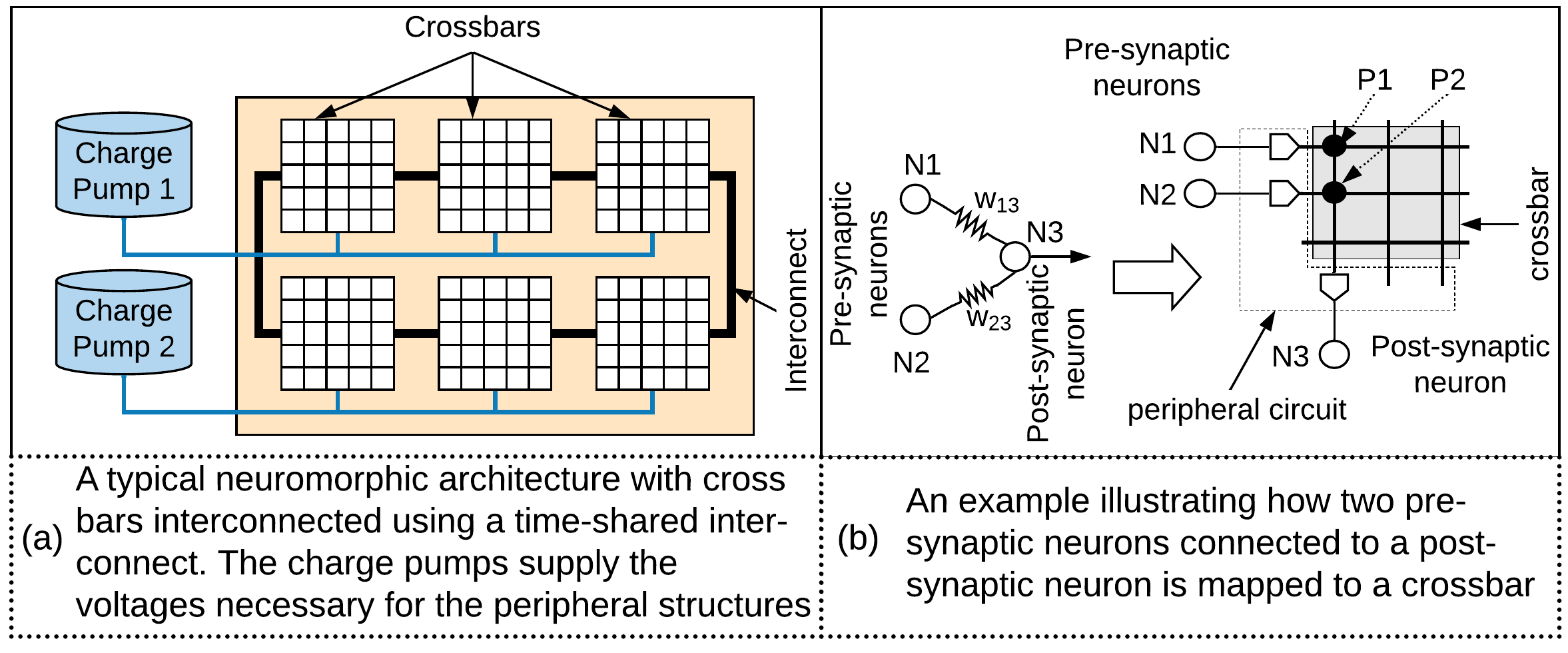}}
	\caption{An illustration of a typical neuromorphic architecture and how SNNs are mapped to a crossbar in this architecture.}
	\vspace{-10pt}
	\label{fig:overview_pcm}
\end{figure}

\mr{\textbf{SNNs} are networks of spiking neurons interconnected via synapses.}
A neuron fires a spike when its membrane voltage exceeds a threshold and subsequently the membrane voltage is reset. The moment of threshold crossing defines the \textit{firing time}. 
\mr{SNNs can be used to implement many machine learning techniques. One example is the supervised approach, where a SNN is first \textit{trained} with examples from the field and then used for \textit{inference} with in-field data.} 
\mr{Performance of supervised machine learning is measured in terms of \textit{accuracy}, which is assessed from inter-spike intervals (ISIs) \cite{grun2010analysis}.} \mr{To define ISI, we let \ineq{\{t_1,t_2,\cdots,t_{K}\}} be a neuron's firing times in the time interval $[0,T]$.} 
The average ISI of this spike train is given by \cite{grun2010analysis}:
\begin{equation}
    \label{eq:isi}
    \footnotesize \mathcal{I} = \sum_{i=2}^K (t_i - t_{i-1})/(K-1).
\end{equation}  
\noindent A \textbf{neuromorphic hardware}, shown in Figure \ref{fig:overview_pcm}(a), consists of 6 crossbars, three of which are connected to charge pump 1 and the remaining three to charge pump 2. All crossbars are interconnected using a time-shared interconnect.
Figure \ref{fig:overview_pcm}(b) illustrates the mapping of an SNN to a crossbar. Synaptic weight \ineq{w_{13}} is programmed on the NVM cell P1 and \ineq{w_{23}} on P2. Output spike voltages \ineq{x_1} from N1 and \ineq{x_2} from N2 inject currents into the crossbar, which are obtained by multiplying a pre-synaptic neuron's output spike voltage with the NVM cell's conductance at the cross-point of the pre- and post-synaptic neurons (following Ohm's law). Current summations along columns are performed in parallel using Kirchhoff’s current law, and implement the sums $\sum_j w_{ij}x_i$, needed for forward propagation of neuron excitation $x_i$.



\begin{figure}[t!]%
    \centering
    \subfloat[Example spike train from N1 of Figure \ref{fig:overview_pcm}(b).]{{\includegraphics[width=9cm]{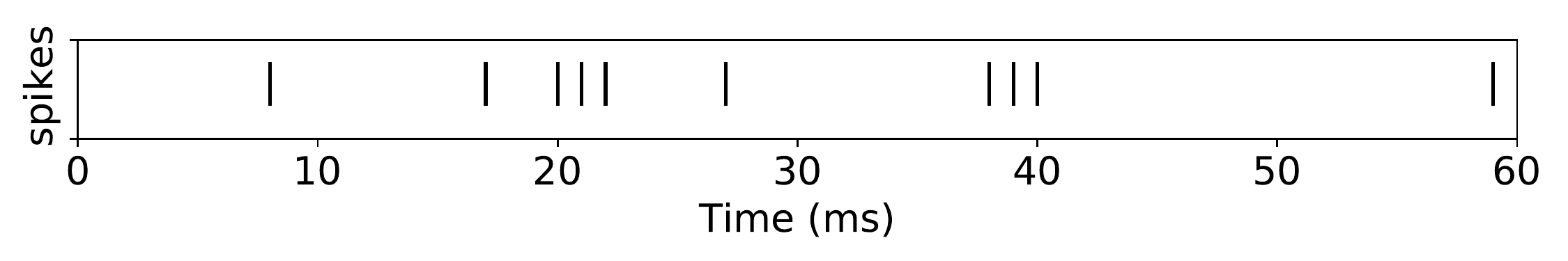} }}%
    \qquad
    \subfloat[Charge pump voltage to process the spike train.]{{\includegraphics[width=9cm]{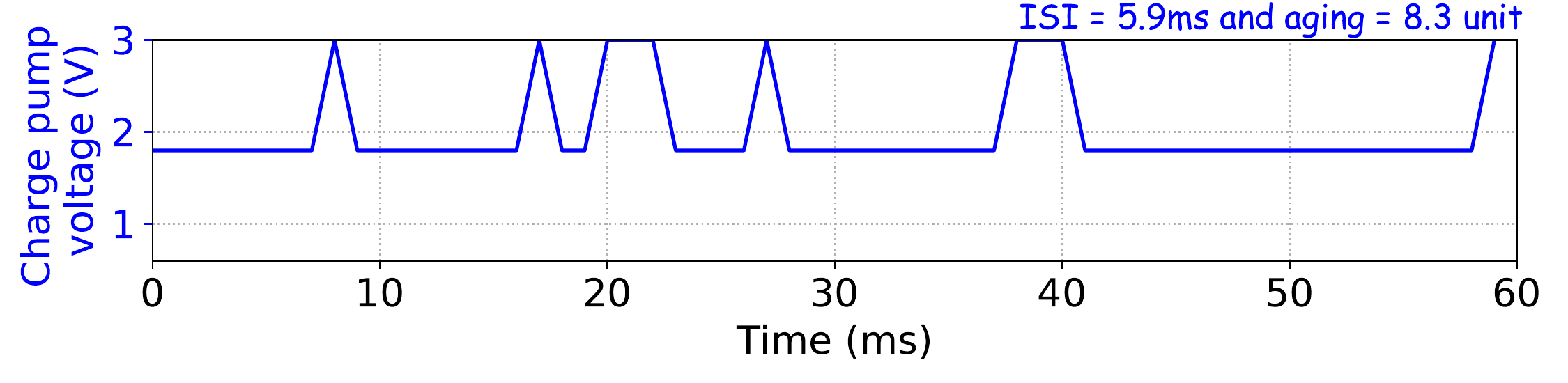} }}%
    \qquad
    \subfloat[Charge pump reset to 1.2V after processing every spike.]{{\includegraphics[width=9cm]{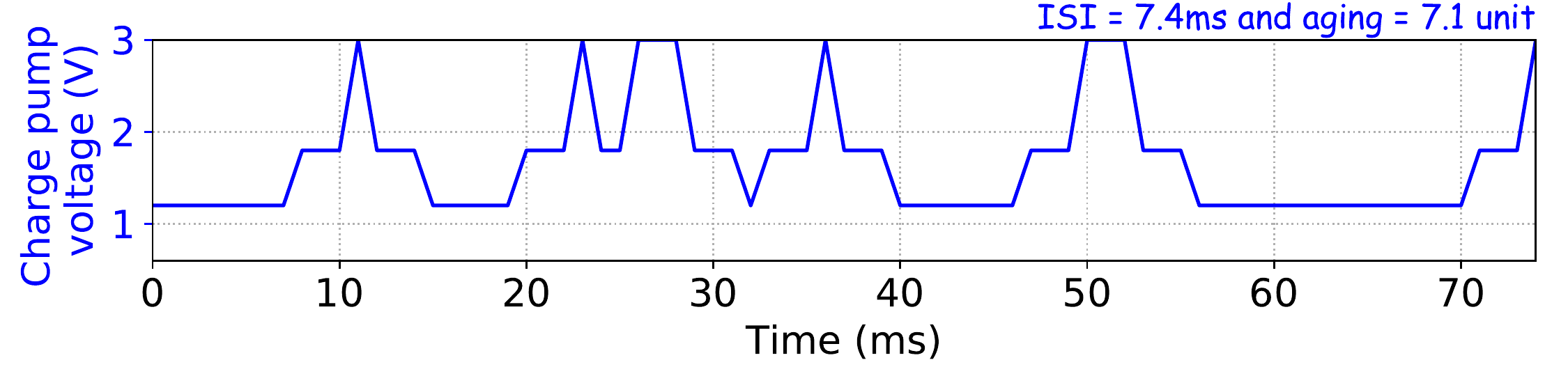} }}%
    \caption{Illustrating the trade-off between charge pump aging and SNN performance, considering PCM crossbars.}%
    \label{fig:cp_motivate}%
\end{figure}

Figure \ref{fig:cp_motivate}(a) shows the spike train generated by N1 of Figure \ref{fig:overview_pcm}(b). Each spike injects current to read the conductance of the NVM cell P1. Figure \ref{fig:cp_motivate}(b) illustrates the charge pump's operating voltage to process this spike train. The charge pump is operated at \ineq{1.8V} for the entire 60ms interval, boosting its voltage to \ineq{3V} only to process spikes. 
Aging of the charge pump is 8.3 units (see Section \ref{sec:reliability} for aging computation) and the average ISI is 5.9ms (Equation \ref{eq:isi}).

Figure \ref{fig:cp_motivate}(c) illustrates the charge pump's operating voltage when it is discharged to \ineq{1.2V} after processing every spike and boosted again to \ineq{1.8V} before processing the next. 
Once discharged, the crossbar becomes unavailable to process spikes, introducing latency in processing the spike train. The average ISI increases to 7.4ms, compared to 5.9ms in Figure \ref{fig:cp_motivate}(b). \textit{ISI deviation leads to accuracy loss} \cite{balaji2019mapping}.
Frequently discharging the charge pump, however, reduces its aging to 7.1 units, compared to 8.3 units in Figure \ref{fig:cp_motivate}(b). This reduction in aging leads to an improvement of mean-time-to-failure (MTTF) of the charge pump by an average 8.7\%. Thus, \textit{aging reduction improves a charge pump's lifetime.} 



\vspace{-10pt}
\section{Proposed Workload-Aware Framework}\label{sec:reliability}
\mr{
We first review NBTI, which is a dominant reliability issue in scaled technology nodes, and then present our proposed framework for PCM-based crossbars. We use characterized NBTI model \cite{gao2017nbti}. Our framework can also be extended with minimal efforts to consider 1) any NBTI model, 2) other NVMs such as FeRAM and Flash, and 3) other reliability issues such as time dependent dielectric breakdown (TDDB), which is still the dominant one in older technology nodes.
}

NBTI aging manifests as 1) decrease in drain current and transconductance, and 2) increase in off current and threshold voltage. NBTI aging is accelerated at high temperature and high oxide electric field. Recent works such as \cite{gao2017nbti} suggest that NBTI is the collective response of two independent defects -- the \textit{as-grown hole traps} (AHTs) and \textit{generated defects} (GDs). AHTs and a small proportion of GDs can be recovered by annealing at high temperatures if the NBTI stress voltage is removed. {We focus on GDs, which contribute to permanent degradation of charge pumps}. In fact, once introduced, GDs cannot be eliminated. Their effect can, however, be delayed by applying lower voltages (i.e., forcefully discharging stressed charge pumps).

To formulate NBTI aging, we divide the SNN execution time \ineq{[0,T]} into \ineq{m} equal intervals \ineq{0 = t_0 < t_1 \cdots<t_m = T}, with \ineq{[t_i,t_{i+1})} as the \ineq{(i+1)^{\text{th}}} interval and \ineq{V_i} is the charge pump's voltage in this interval. Reliability at the end of SNN execution can be expressed as \ineq{R(T) = e^{-\left(\sum_{i=0}^{m-1}G(V_i)\right)^{\beta}}},
where 
\ineq{G(V_i)} is the generated defect at voltage \ineq{V_i}, expressed as power law, \ineq{G(V_i) = g_0\cdot(V_i-V_\text{th})^m\cdot(t_{i+1}-t_i)^n} and \ineq{\beta,g_0,m,n} are material-dependent constants~\cite{gao2017nbti}.
We define NBTI aging in a stressed charge pump as
\begin{equation}
\label{eq:eq11}
\scriptsize \mathcal{A} = \sum_{i=0}^{m-1}g_0\cdot(V_i-V_\text{th})^m\cdot(t_{i+1}-t_i)^n, \text{ such that } R(T) = e^{-\mathcal{A}^{\beta}}.
\end{equation}

\noindent Here \eqref{eq:eq11} assumes all synapses are mapped to the same crossbar, which is connected to a single charge pump. \mr{In practice, however, 1) synapses are distributed across different crossbars because a crossbar can accommodate only a limited number of synapses and 2) a neuromorphic hardware typically has more than one charge pump to limit the power supply load.} We now describe how to extend \eqref{eq:eq11} to incorporate these practical constraints. 

We consider the SNN \ineq{\mathcal{G}}, with \ineq{N} neurons and \ineq{S} synapses, excited with an input over the time interval \ineq{[0,T]}. 
We arrange the spikes in this interval by synapses they excite as
\begin{equation}
    \label{eq:spike_time_synapse}
    \footnotesize \mathcal{S} = \{\tau_1^1,\tau_2^1,\cdots,\tau_{k_1}^1\},\{\tau_1^2,\tau_2^2,\cdots,\tau_{k_2}^2\},\cdots,\{\tau_1^S,\tau_2^S,\cdots,\tau_{k_S}^S\}, 
\end{equation}
where \ineq{\tau_j^s} is the \ineq{j^\text{th}} spike on \ineq{s^\text{th}} synapse of the SNN. We introduce the following notation.

\vspace{-10pt}
\begin{footnotesize}
\begin{align*}
    \mathcal{A}_{s} & \text{: aging to process spike train } \{\tau_1^s,\cdots,\tau_{k_s}^s\} \text{ on } s^\text{th} \text{ synapse}\\
    C & \text{: number of crossbars}\\
    L & \text{: number of charge pumps}\\
    \mathcal{M} \in \mathbb{R}^{S\times C} & \text{:synapse-to-crossbar mapping, such that}
\end{align*}
\end{footnotesize}
\vspace{-10pt}
\begin{equation}
    \label{eq:mapping}
    \footnotesize m_{ij}\in \mathcal{M} = \begin{cases}
    1 & \text{if synapse } i \text{ is mapped to crossbar } j\\
    0 & \text{otherwise}
    \end{cases}
\end{equation}
\vspace{-10pt}
\begin{footnotesize}
\begin{align*}
    \mathcal{P}\in \mathbb{R}^{C\times L} & \text{:crossbar-to-charge pump mapping, such that}
\end{align*}
\end{footnotesize}
\vspace{-10pt}
\begin{equation}
    \label{eq:placement}
    \footnotesize p_{jk}\in \mathcal{P} = \begin{cases}
    1 & \text{if crossbar } j \text{ is powered by charge pump } k\\
    0 & \text{otherwise}
    \end{cases}
\end{equation}
Combining these two equations, we generate the synapse-to-charge pump mapping as
\begin{equation}
    \label{eq:placement}
    \footnotesize m_{ij}\cdot p_{jk} = \begin{cases}
    1 & \text{if synapse } i \text{ is powered by charge pump } k\\
    0 & \text{otherwise}
    \end{cases}
\end{equation}

The total aging of charge pump \ineq{k} is therefore
\begin{equation}
    \label{eq:wear_out_cp}
    \footnotesize \text{aging}_k = \sum_{i=1}^S \sum_{j=1}^C m_{ij}\cdot p_{jk}\cdot \mathcal{A}_i
\end{equation}

\begin{figure}[t!]
	\centering
	\centerline{\includegraphics[width=0.99\columnwidth]{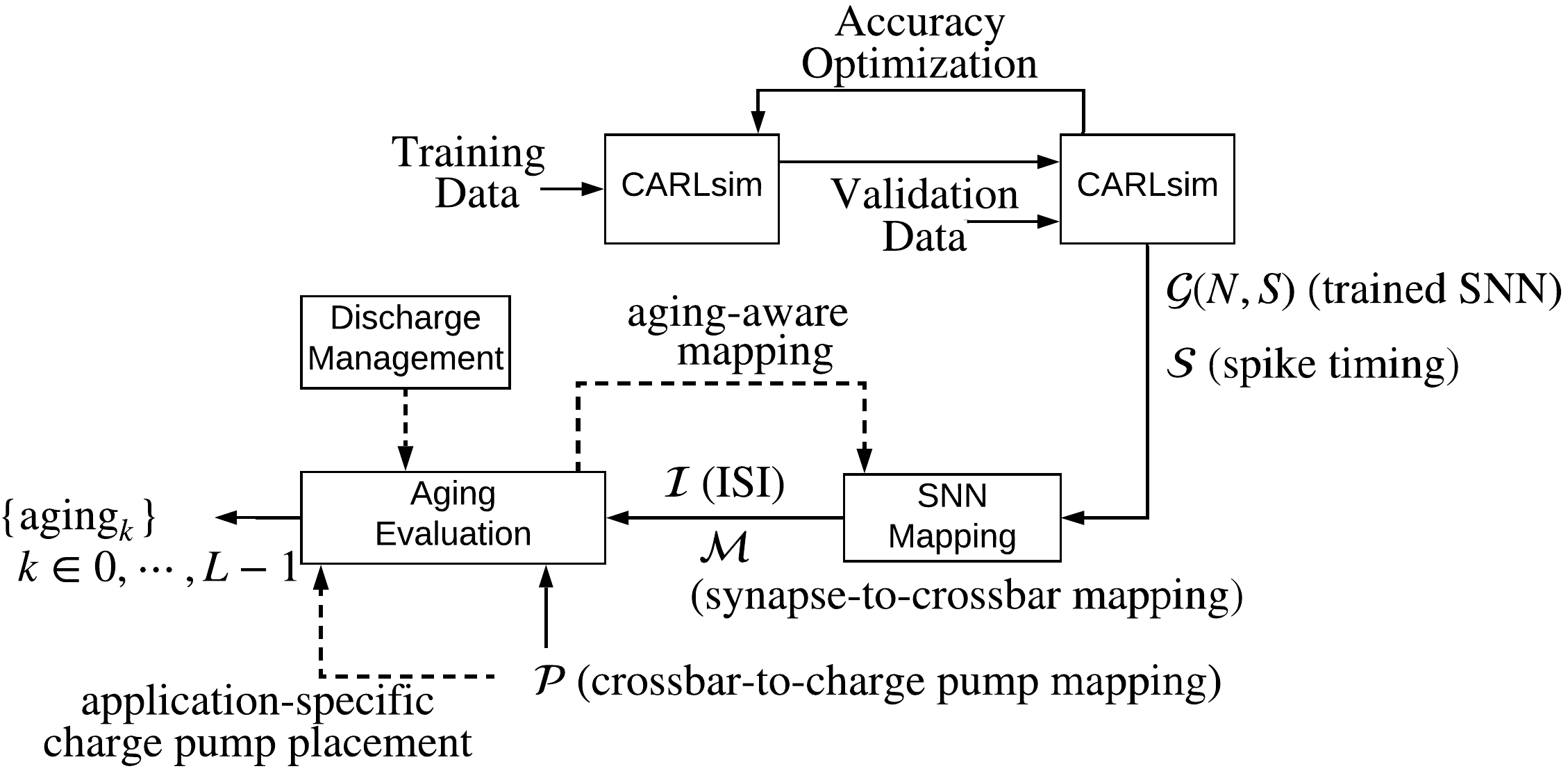}}
	\caption{Framework to evaluate aging of charge pumps.}
	\vspace{-10pt}
	\label{fig:framework}
\end{figure}

\noindent \textbf{Proposed Framework --} Figure \ref{fig:framework} illustrates our framework to evaluate aging of charge pumps in a neuromorphic hardware.
We use CARLsim \cite{Chou2018CARLsim4} to train SNN models.
The output of CARLsim are the trained weights and the precise times of spikes on all synapses of the SNN \ineq{\mathcal{S}}. A SNN mapping approach such as \cite{balaji2019mapping} uses CARLsim output to generate a synapse-to-crossbar mapping \ineq{\mathcal{M}}, optimizing some objective function. In \cite{balaji2019mapping}, the objective function is to minimize the number of spikes communicated between crossbars, which leads to lower energy and latency on the shared interconnect. Once the SNN is mapped to crossbars of the hardware, its performance is obtained in terms of the inter-spike interval \ineq{\mathcal{I}} using \eqref{eq:isi}.
Using this synapse-to-crossbar and crossbar-to-charge pump mapping, our novel formulation in \eqref{eq:wear_out_cp} evaluates the aging of all charge pumps in the hardware when executing an SNN workload. This design flow is shown using solid arrows.

Figure \ref{fig:framework} also illustrates three future directions based on this framework using dashed arrows.
First, \emph{Aging Evaluation}, as developed in \eqref{eq:wear_out_cp}, can be combined with the \emph{SNN Mapping} step to generate an optimum mapping of the SNN to the hardware that balances aging of all charge pumps. This is shown by the dashed arrow labeled \emph{aging-aware mapping}. Second, crossbar-to-charge pump mapping can be optimized to achieve a desired lifetime of charge pumps for executing the SNN. This is shown using the dashed arrow labeled \emph{application-specific charge pump placement}. Third, strategies can be developed to discharge charge pumps at run-time, improving their lifetime. This is shown in the \emph{Discharge Management} step. 

\vspace{-10pt}
\section{Evaluation Results}\label{sec:evaluation}
This section presents evaluation results using our framework. We use the neuromorphic hardware of Figure \ref{fig:overview_pcm}(a) to evaluate the following SNNs \cite{mlperf,Das2018HeartbeatECG,Das2018UnsupervisedReadout,balaji2019mapping}. 

\begin{table}[h!]
	\renewcommand{\arraystretch}{0.8}
	\setlength{\tabcolsep}{2pt}
	\centering
	\begin{threeparttable}
	{\fontsize{6}{10}\selectfont
		\begin{tabular}{c|cl|c}
			\hline
			\textbf{SNN} & \textbf{Synapses} & \textbf{Topology} & \textbf{Spikes}\\
			\hline
			ImgSmooth & 136,314 & FeedForward (4096, 1024) & 17,600\\
			EdgeDet & 272,628 &  FeedForward (4096, 1024, 1024, 1024) & 22,780\\
			MLP-MNIST  & 79,400 & FeedForward (784, 100, 10) & 2,395,300\\
			HeartEstm & 636,578 & Recurrent & 3,002,223\\
			HeartClass  & 2,396,521 & CNN\tnote{1} & 1,036,485\\
			CNN-MNIST & 159,553 & CNN\tnote{2} & 97,585\\
			LeNet-MNIST & 1,029,286 & CNN\tnote{3} & 165,997\\
			LeNet-CIFAR  & 2,136,560 & CNN\tnote{4} & 589,953 \\
			\hline
	\end{tabular}}
	\begin{tablenotes}\scriptsize
        \item[1.] (82x82) - [Conv, Pool]*16 - [Conv, Pool]*16 - FC*256 - FC*6
        \item[2.] (24x24) - [Conv, Pool]*16 - FC*150 - FC*10
        \item[3.] (32x32) - [Conv, Pool]*6 - [Conv, Pool]*16 - Conv*120 - FC*84
        \item[4.] (32x32x3) - [Conv, Pool]*6 - [Conv, Pool]*6 - FC*84 - FC*10
    \end{tablenotes}
	\end{threeparttable}
	\vspace{-10pt}
\end{table}
\vspace{-10pt}

\subsection{Evaluating reliability of SNN mapping strategies}
\begin{figure}[t!]
	\centering
	\centerline{\includegraphics[width=0.99\columnwidth]{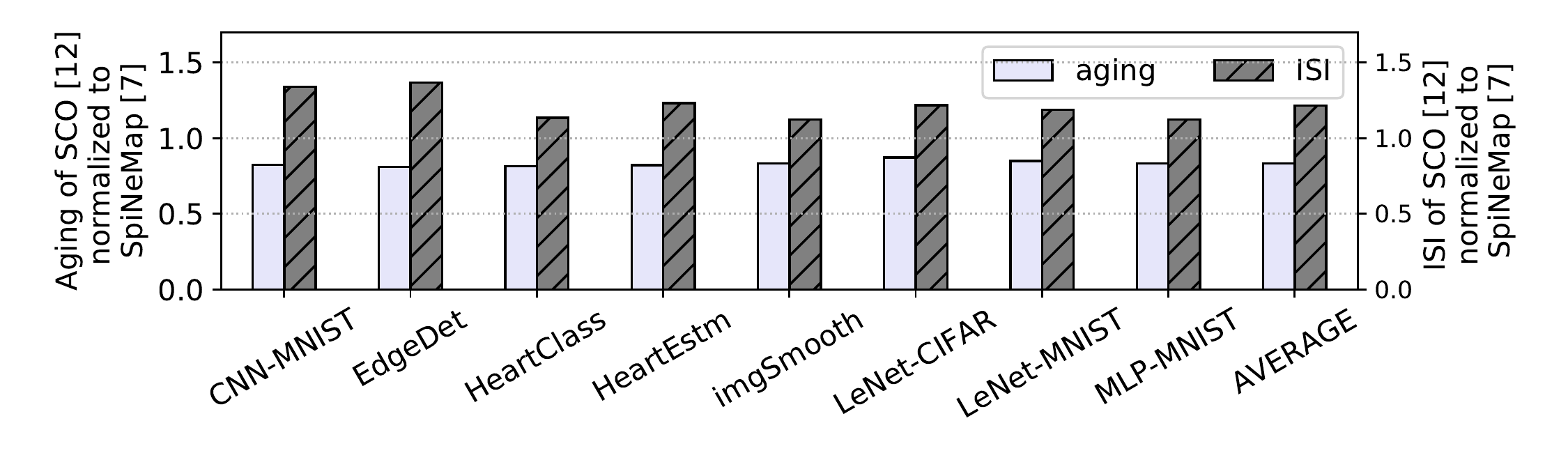}}
	\caption{Aging and ISI of SCO \cite{lee2019system} vs. SpiNeMap \cite{balaji2019mapping}.}
	\vspace{-10pt}
	\label{fig:tradeoff}
\end{figure}

\mr{
We use our framework to evaluate two state-of-the-art SNN mapping strategies -- SCO \cite{lee2019system} and SpiNeMap \cite{balaji2019mapping}, in terms of performance (measured as change in ISI) and reliability (measured as aging). Figure \ref{fig:tradeoff} illustrates the result of SCO, normalized to SpiNeMap.
SCO, which balances crossbar utilization, has on average 16.4\% lower aging (better lifetime) than SpiNeMap for these workloads. This is because SpiNeMap explicitly minimizes spike latency on the shared interconnect. To do so, some crossbars get more utilized than others. Heavily utilized crossbars activate charge pumps more frequently, causing their higher aging. Conversely, SpiNeMap has lower ISI change (higher performance). SCO has on average 21\% higher change in ISI than SpiNeMap.
\textit{From a performance perspective, SpiNeMap is better than SCO, while from a reliability perspective, SCO is better than SpiNeMap.}
}


\vspace{-15pt}
\subsection{Discharging stressed charge pumps}
\begin{figure}[t!]%
    \centering
    \subfloat[Aging for different discharge intervals normalized to the aging when charge pumps are not discharged.]{{\includegraphics[width=9cm]{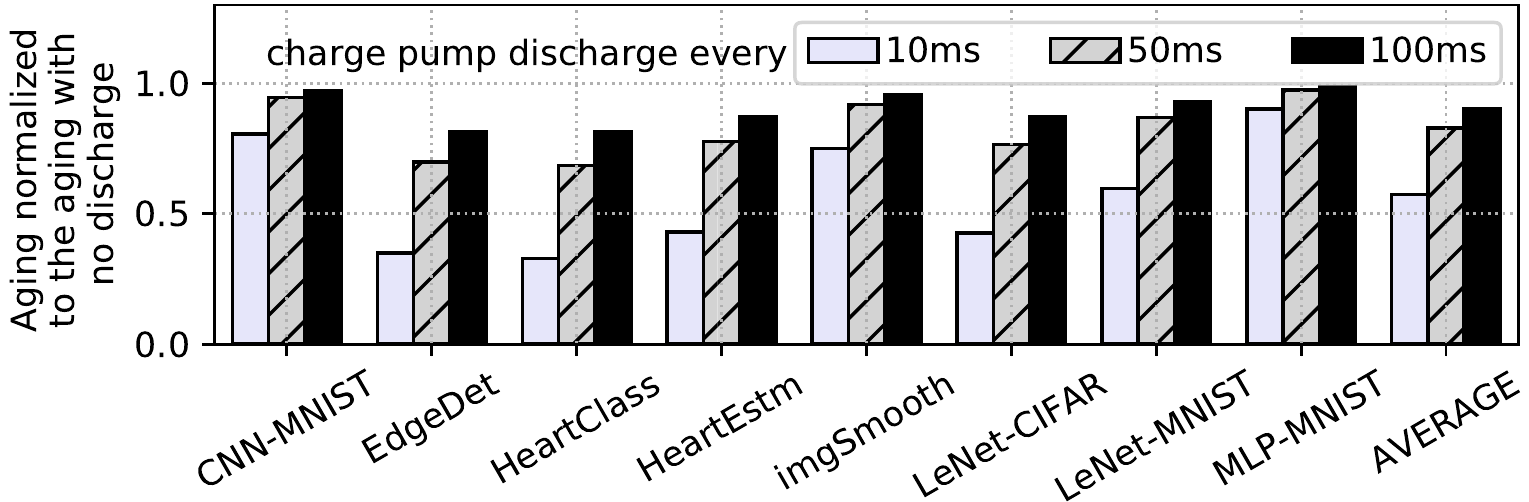} }}%
    \qquad
    \subfloat[ISI for different discharge intervals normalized to the ISI when charge pumps are not discharged.]{{\includegraphics[width=9cm]{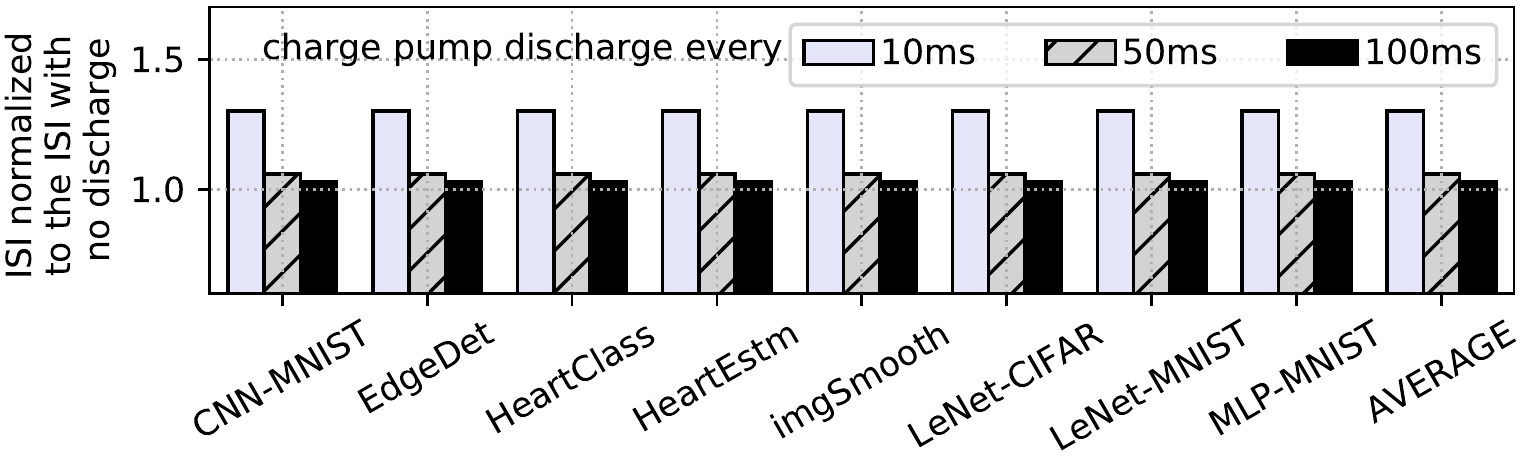} }}%
    \caption{Aging and ISI with different discharge intervals.}%
    \vspace{-12pt}
    \label{fig:discharge}%
\end{figure}

Figure \ref{fig:discharge} illustrates aging and ISI with discharge intervals of 10ms, 50ms, and 100ms for the evaluated SNN workloads, normalized to when charge pumps are stressed for the entire execution duration. We make the following three key observations. 
First, aging is the lowest for discharge interval of 10ms, while ISI variation is the highest.
\mr{
This is because, with smaller discharge intervals, a charge pump's internal CMOS devices recover partially from stress and therefore, the rate of aging reduces improving lifetime. The performance is lower because of the delay introduced in frequent charge pump discharge.
}
Second, when the discharge interval changes from 10ms to 100ms, aging increases, reducing charge pump's lifetime, and ISI variation reduces, improving application performance. Third, aging of charge pumps varies across different SNN workloads. \mr{For MLP-MNIST, aging increases by 10\% when the discharge interval increases from 10ms to 100ms, while for LeNet-CIFAR, aging increases by a factor of 2 for the same range. This is because for MLP-MNIST, spikes are generated less frequently due to sparsity of synaptic weights. There is therefore, no significant variation in aging when charge pumps are discharged differently. The ISI variations are, however, due to delay of spike propagation when charge pumps are being discharged. We see no significant variations across different workloads.} Our framework enables exploration of SNN workload-specific lifetime and performance trade-offs.

\vspace{-10pt}
\section{Discussion and Future Outlook}\label{sec:conclusion}
\mr{
Aging-related defects in charge pumps constitute a critical bottleneck to the prolonged operating lifetime of neuromorphic hardware.
}
These defects are {different} from an NVM cell's endurance failures, which are due to {repeated} programming of the cell. 
\mr{
In recent prototypes, e.g. \cite{song2018high}, PCM endurance is in the order of \ineq{10^{7}} cycles (\ineq{\approx} 4-5 years lifetime). A charge pump's lifetime is \ineq{\approx} 2-3 years operating at 3V supply.
}
Impact-wise, aging issues in a neuromorphic hardware arise during inference (reading of synaptic weights) and training (update of synaptic weights) in supervised machine learning, while endurance issues arise only during training.

To this end, we proposed a novel framework to evaluate SNN workload-specific lifetime and performance trade-offs in neuromorphic architectures. 
The framework incorporates the CARLsim simulator to extract the precise time of spike generation on all synapses of an SNN workload. Using this timing information, together with 1) synapse-to-crossbar mapping, and 2) crossbar-to-charge pump mapping, this framework evaluates aging of different charge pumps when executing an SNN workload. 
\mr{We use this framework to evaluate two state-of-the-art SNN mapping strategies in terms of performance and reliability.
} 
We also demonstrated lifetime and performance trade-offs by changing the charge pump's discharge interval.
Our framework can also incorporate: \textbf{1) other SNN simulators} such as Brian \cite{goodman2009brian}, and \textbf{2) other reliability issues} such as electromigration \cite{DasCASES}.





\vspace{-10pt}

\section*{Acknowledgment}
This work is supported by the National Science Foundation Award CCF-1937419 (RTML: Small: Design of System Software to Facilitate Real-Time Neuromorphic Computing).
\vspace{-10pt}

\bibliographystyle{IEEEtran}
\bibliography{IEEEabrv,neuromorphic_computing}
\end{document}